\documentstyle[preprint,aps,eqsecnum]{revtex}

\begin{document}
\draft
\preprint{gr-qc/9304010  Alberta-Thy-2-93}

\title{General covariance,  loops, and matter}

\author{Viqar Husain}

\address{Theoretical Physics Institute, University of Alberta\\
Edmonton, Alberta T6G 2J1, Canada.}

 \maketitle

\begin{abstract}
A four dimensional generally covariant field theory is presented
which
describes
non-dynamical  three geometries coupled to scalar fields.

The theory has an  infinite
number of physical observables (or constants of the motion) which are
constructed from
loops made from  scalar field configurations.  The Poisson algebra of
these observables
is closed and is the same as that for the 3+1 gravity loop variables
in the Ashtekar
formalism. The theory also has observables that give the areas of
open
surfaces
and the volumes of finite regions.

Solutions to all the Hamilton-Jacobi equations for the theory and the
Dirac
quantization
conditions in the coordinate representation are given.  These
solutions are
holonomies
based on matter loops.
A brief discussion of the loop space representation for the quantum
theory
is also given
together with some implications for the quantization of 3+1 gravity.
\end{abstract}
\pacs{04.60.+n,04.20.Fy}
\vfill
\eject

\section{Introduction}
The Ashtekar Hamiltonian variables \cite{ash1} for general relativity
and the
associated
loop variable
approach \cite{lc1} to quantum gravity have inspired a number of toy
models
that have
similar phase
space   variables and are amenable to quantization via the loop
representation.
 The goal
  of these models is to try and mimic the basic  features of  the
constraint
structure of general
  relativity in the hope of gaining some insights into quantization.
Most of
the models
  considered so far have not included matter fields and
  it is not clear  what role such fields will play in the loop
variable
methods.  It is
  clearly important to include matter in this approach to quantum
gravity and
here we
  consider  a model that  attempts to do this.

 A seperate but essential reason for including matter fields is the
issue
 of  {\it physical observables}, which are phase space functions
which have
  weakly vanishing poisson brackets with all the constraints.  For
the case of
  pure (spatially closed) gravity, no such observables are known but
there
  are indications that inclusion of matter to define reference
   systems may help to determine physical observables
\cite{dw,rov,kuchtor}.

   These observables play  the essential role in
  one approach to the quantization of generally covariant theories.
In this
method
  the goal is to define the quantum theory by finding a
representation
   of the Poisson algebra of  a complete set of  observables.  This
has been
applied
   successfully to  2+1 gravity \cite{3grav} but  it cannot  even be
started
for 3+1
gravity since no
   such observables are known (for the compact case).  There is some
recent
work
   indicating that  the natural observables  in general relativity
are
associated with
   functions that measure area or volume\cite{ars}.  These are
kinematical
constructions and
   are not invariant under the dynamics generated via the Hamiltonian
constraint. They are
    also not naturally spatial diffeomorphism  invariant since
auxiliary
variables are used
    in some of the   definitions.  One can attempt to partly rectify
this
   situation by the inclusion of matter fields, and use them in place
of the
auxiliary
   variables, to make truly diffeomorphism invariant  observables.
Attempts in
this
   direction have been made recently using scalar fields \cite{csurf}
and
anti-symmetric
tensor
   fields \cite{lsurf} to define area observables, and their spectra
given via
a
quantization
   based on the loop representation.
   The present  work is also a partial attempt in this direction but
the
emphasis is on
   constructing diffeomorphism invariant loop observables.

  In the case of general relativity without matter fields, the
Rovelli-Smolin
loop
 observables, which are based on holonomies, play an
 important role in the construction of the loop representation. These
variables
 are
 not diffeomorphism invariant (since they are functions of the
auxiliary loops
as well),
  but nevertheless a representation of their Poisson algebra
provides an
important step
 towards  the solution of  the diffeomorphism constraint.

 In this paper we discuss a model that has  spatially diffeomorphism
invariant
loop
 observables where the loops are made of matter. A four dimensional
generally
covariant model is discussed in the next section and its Hamiltonian
version
and  physical observables are given.  Section 3 is a discussion of
3+1 gravity
with
scalar fields,  which is obtained from this model by including the
Hamiltonian
constraint
on the phase space. There is also a discussion of the
 the Hamilton-Jacobi equations for the
constraints, which are completely solved for the model and
  partially solved for 3+1 gravity by
using a holonomy functional. Section 4 is a brief discussion of
the Dirac and loop quantizations of the model followed by a summary
and discussion in section 5.

\section {Generally covariant  model and observables}
   There is a type of generally covariant field theory that is not
topological,
has  internal Yang-Mills symmetry, but no dynamics \cite{hk}.  The
lagrangian
is a four form constructed from the curvature of an so(3)  gauge
field and
a dreibein:
\begin{equation}
 S_1=\int_M d^4x \epsilon^{\alpha\beta\gamma\delta} \epsilon_{ijk}
e_\alpha^i e_\beta^j F_{\gamma\delta}^k
\end{equation}
 The Hamiltonian formulation of this  model is similar to that
for general relativity in the Ashtekar formulation except that the
Hamiltonian
constraint
is absent (or  equivalently, vanishes strongly): the only first class
 constraints that appear on the phase space correspond to spatial
diffeomorphisms
  and internal gauge transformations \cite{hk}.  This theory
therefore  has
three
local degrees
of freedom and is
effectively a three dimensional  theory since the dynamics is
trivial.  There
are indications
that it is related to a c=0 limit of general relativity \cite{csurf}.
There is
a
similar model in three
dimensions that may be viewed as arising from the hamiltonian
decomposition of
a 2D
de-Sitter group  Chern-Simons theory \cite{h}.

 Here we consider theories of this type with coupling to a number of
scalar
fields.  Since
 the metric obtained from the dreibein
$g_{\alpha\beta}=e_\alpha^ie_\beta^i$ is
degenerate,
 the usual  form of coupling scalar fields is not possible.  However
there is a
 way to couple {\it non-dynamical} scalar fields (which is in the
spirit of the
model (2.1))
 using the vector density
  $n^\alpha \equiv \epsilon^{\alpha\beta\gamma\delta}e_\beta^i
e_\gamma^j
e_\delta^k
  \epsilon_{ijk}$.  We add to  $S_1$ the term
\begin{equation}
    S_2 = \int_M d^4x n^\alpha \pi_n\partial_\alpha\phi_n
\end{equation}
where $\pi_n$ and $\phi_n$ are a pair of scalar fields for
$n=1,...,N$. The
equations
of motion for these scalars give $n^\alpha \partial_\alpha\phi_n=0$
and
$ \partial_\alpha(\pi_n n^\alpha)=0$ which effectively imply that
the fields
do not evolve
(since $n^\alpha$ is the degeneracy direction for $g_{\alpha\beta}).$

The Hamiltonian theory for the action $S=S_1+S_2$ on $M=\Sigma\times
R$ is
easily
 obtained since the action is in first order form. The 3+1 form of
the
 action is
\begin{equation}
 S=\int_R dt\int_\Sigma d^3x \Bigl[ E^{ai}\dot{A}_a^i +
\Pi_n\dot{\phi}_n
  + N^a( F_{ab} E^{bi}  +  \Pi_n\partial_a \phi_n)
   +A_0^i(D_aE^{ai})\Bigr]
 \end{equation}
   where $N^a=e^{ai}e_0^i$ and $e^{ai}$ is the inverse of the
  projection of the  dreibein onto the surface $\Sigma$
  (where it is invertible).  The
canonical phase space variables are those  of so(3) Yang-Mills theory
with
conjugate variables $(A_a^i, E^{ai})$, together with the   $n$ scalar
field
variables
 $(\phi_n, \Pi_n)$.  $E^{ai}\equiv
\epsilon^{abc}\epsilon^{ijk}e_b^je_c^k$ and
$\Pi_n\equiv n^0\pi_n$. The first  class constraints on the phase
space are the
Gauss law and  spatial diffeomorphisms obtained by varying $S$ with
respect to $A_0^i$ and $N^a$.
\begin{eqnarray}
G^i& = &D_a E^{ai} = 0, \\
 C_a& = &F_{ab} E^{bi}  +  \Pi_n\partial_a \phi_n = 0
 \end{eqnarray}
where $D_a$ and $F_{ab}$ are the covariant derivative and curvature
of $A$.
Since
there is no Hamiltonian constraint, the `dynamics' generated by a
linear
combination
of the constraints is pure gauge.
 This completes the description of the Hamiltonian theory. (The basic
 reason that the Hamiltonian constraint is absent in the theory
  is a result of the presence of a degeneracy direction given
  by the vector field density $n^\alpha$.
  When converted into a vector field by means of an auxiliary
  foliation of $M$, it Lie derives the metric \cite{hk}).

  This Hamiltonian  theory for one scalar field has been considered
in a
context related
  to extracting
   area observables as constants of the
 motion and their  quantization \cite{csurf}. Here we give a
discussion of loop
observables in this
 theory,  and point out that the Gauss law  invariant loop
observables
introduced on the
 Ashtekar  phase  space of  general relativity  can be made
diffeomorphism
invariant
  as well when scalar fields are present.

To define loops on the 3D surfaces  $\Sigma$ we need two scalar
fields,
 and a loop
 $\gamma[\phi_1,\phi_2]$ is
 defined as the intersection of two surfaces $\phi_1 = c_1,
\phi_2=c_2$. A
 vector density tangent to the loop is
    \begin{equation}
     \gamma^a =\epsilon^{abc}\partial_b\phi_1\partial_c \phi_2
   |_{\phi_1=c_1,\phi_2 = c_2}
   \end{equation}
 This form is not necessary for any computations, for which we need
only the variation of $\phi_n=c_n$:
  $\delta\phi_n +\delta \gamma^a\partial_a\phi_n =0$.
  (Regarding this method we note that in some recent work
  \cite{newc} Newman and Rovelli have used scalar fields
  in a similar way to
  solve (classically) the Gauss law constraints in Yang-Mills theory.
  In particular, for the Abelian theory they set the electric field
  $E^a := \epsilon^{abc}\partial_bu\partial_cv$ for two scalar fields
 $(u,v)$, which solves the Gauss law. The electric field lines in
this solution are tangent to the loops defined by $u=c_1,v=c_2$).

 The first few loop observables are  defined as
\begin{eqnarray}
  T^0[\phi_1,\phi_2,A](c_1,c_2)& = &
TrPexp\int_{\gamma[\phi_1,\phi_2]}ds\dot{\gamma}^a(s)
  A_a(\gamma(s)) \\
   T^1[A,E,\phi_1,\phi_2](c_1,c_2)& = &\int_{\gamma[\phi_1,\phi_2]}
ds w_a(s)
    Tr[ E^a(\gamma(s)) U_\gamma(s,s)] \\
 T^2[A,E,\phi_1,\phi_2](c_1,c_2)& = &\int_{\gamma[\phi_1\phi_2]} ds
\int_{\gamma[\phi_1,\phi_2]} dt
w_a(s) w_b(t) \nonumber \\
&  & Tr[ E^a(\gamma(s ))U_\gamma(s,t)E^b(\gamma(t))U_\gamma(t,s)]
\end{eqnarray}
where the 1-form  density
$$w_a\equiv \epsilon_{abc}
 \dot{\gamma}^b {\delta\gamma^c\over \delta \phi_1}.
  $$
 These are functionals of the fields $(A,E,\phi_1,\phi_2)$ and
 functions of the two parameters $(c_1,c_2)$.
All the other observables $T^N$, with $N$ $E$-insertions in the
holonomies are constructed  in a similar way.
These observables are modeled after the (Gauss law invariant)
Rovelli-Smolin loop variables \cite{lc1} and they have essentially
the
same closed Poisson algebra since they are independent of the scalar
 field momenta $\Pi_n$.
 The novelty here is that the elements of this $T$ algebra  are
{\it the constants of the motion} associated
with the action $S$.  It is clear by inspection that these
 observables are diffeomorphism invariant (and this may be checked by
 explicitly  computing Poisson brackets with $C_a$).

There are an infinite number of loops determined by the
configurations of the two scalar fields and the constants $c_1,c_2$.
It is also easy to visualize
how multiloops may arise by considering, for example,  the case where
one of the scalar fields defines a plane and the other defines a
nearly
parallel plane but with a large number of `bumps' that intersect
the first plane.

 These $T^N$ physical observables are however not the complete set
 since they do not involve the scalar field momenta $\Pi_n$.
 The diffeomorphism
invariant observables involving  these are,
for $n=1,2$, $ P_n[\Pi_n] =  \int_\Sigma d^3x \Pi_n $.
These have non-trivial Poisson brackets with the $T^N$. For example
\begin{eqnarray}
\{T^0, P_n\}& = &\int_\Sigma d^3x {\delta T^0\over
 \delta \phi_n} \nonumber \\
 & = &\int_{\gamma[\phi_1,\phi_2]} ds \epsilon^{abc}w_c(s)
    Tr[ F_{ab}(\gamma(s)) U_\gamma(s,s)].
\end{eqnarray}
In general $\{T^N, P_n\} = \int_\Sigma d^3x
(\delta T^N/  \delta \phi_n)(x)$, and the functional
 derivative with respect to $\phi_n$ acts effectively
 to shift the loop $\gamma^a[\phi_1,\phi_2](c_1,c_2)$ by shifting
  $\phi_1,\phi_2$ but leaving $c_1,c_2$ fixed:
 $\delta/\delta \phi_1 =
 (\delta \gamma^a/\delta \phi_1)\delta/\delta \gamma^a$.
The Poisson brackets of any observable involving the
 momenta $\Pi_n$ will therefore not close with the $T^N$.
To obtain closure it appears that the set
of $T^N$ will have to be extended to include all the additional
loop observables that involve higher functional derivatives
 of the form $\delta w/\delta \phi_n$ in the integrands of the $T^N$.
 Such terms result from
 calculating the Poisson brackets. This  extension can be
  done but the resulting algebra loses the elegance of just the
 $T^N$ observables.

We note that there are other diffeomorphism invariant observables
that may be constructed from scalar fields. These are the area
observables
discussed
by Rovelli \cite{csurf} which are defined using one scalar field:
\begin{equation}
 A=\int_{S[\phi]} d^2\sigma \sqrt{h}\equiv
\int_{S[\phi]} d^2\sigma \sqrt{E^{ai}E^{bi}n_an_b}
\end{equation}
where $n_a=\epsilon_{abc}(\partial x^b/\partial\sigma^1)(\partial
x^c/\partial\sigma^2)$.
 This observable commutes  with $T^0$:
 \begin{eqnarray}
 \{T^0[\phi_1,\phi_2,A],A[\phi_1,E]\}& = &
 \int_{\gamma[\phi_1,\phi_2]}ds\int_{S[\phi_1]}d^2\sigma
\delta^3(\gamma(s),\sigma) \nonumber \\
 &  &
 \dot{\gamma}^a Tr[E^bU](\gamma(s)) {n_an_b\over \sqrt{h}}(\sigma)
 \nonumber \\
& = &0 \end{eqnarray}
since the normals $n_a$ to the surfaces $\phi_1=C$ are perpendicular
to the
tangent vector to the loop.  A further observation regarding $A$ is
that, since
we have two scalar fields,
 it may be defined  for {\it open surfaces} where the boundary of the
 surface is
 determined by the loop $\gamma[\phi_1,\phi_2]$. This a different way
of
 constructing observables associated with open surfaces than the one
 given by Smolin \cite{lsurf} where an Abelian gauge field is used to
specify
the
 boundary of a surface specified by an antisymmetric tensor field.

Another diffeomorphism invariant observable is
\begin{equation}
 Q[E,\phi] = \int_\Sigma d^3x
\sqrt{E^{ai}E^{bi}\partial_a\phi_1\partial_b\phi_1}.
\end{equation}
This functional is essentially the  same as the one used  to define
the `weaves', which are distributional dreibeins taking non-zero
values only on
given
configurations
of loops.  It is shown in ref.\cite{ars} that this functional may be
converted
into a
well defined
 operator in the loop space representation, and that its eigenstates
are
distributional dreibeins
  which take values on sets of loops.
  Whereas in \cite{ars} this observable is defined using an auxiliary
1-form
$\omega_a$
on $\Sigma$,
we see that with a scalar field one can define it using
$\partial_a\phi$
thereby
 converting it into a diffeomorphism invariant phase space
functional.
  (We also note that the integrand is the square root of  the
scalar field contribution to the Hamiltonian constraint for gravity
\cite{matter}. See
eqn. (3.1) below).
 As for the area observable, $Q$ commutes with $T^0$
  for the same reason, namely $\dot{\gamma}^a\partial_a \phi_1 =0$ on
the loop:
 \begin{eqnarray}
 \{Q[A,\phi_1],T^0[A,\phi_1,\phi_2]\}& = &
  \int_{\gamma[\phi_1,\phi_2]}ds
 {\dot{\gamma}^a\partial_a\phi_1 \partial_b\phi_1Tr[E^bU]
 \over \sqrt{E^{ai}E^{bi}\partial_a\phi_1\partial_b\phi_1}}
  \nonumber \\
  & = & 0
 \end{eqnarray}

In fact this result generalizes: all the $T^N$ defined above commute
with $A$
and $Q$.
Thus it is possible to specify sets of three mutually commuting
observables
that are associated with loops, areas and metrics: any one of the
$T^N$,   $A$
 and $Q$.
It may be possible  to find a representation where the corresponding
operators
 have simultaneous eigenstates.

A final diffeomorphism invariant observable is one that measures the
spatial
volume:
 \begin{equation}
 V = \int d^3x \sqrt{det E(x)}.
 \end{equation}
     Since its definition requires no auxiliary fields
it is diffeomorphism invariant without the scalar fields. However,
the two
scalar
 fields allow one to define diffeomorphism invariant boundaries of a
spatial
 region, and so it becomes possible to limit the range of  volume
integration
to regions
 bounded by the surfaces $\phi_1=c_1$, $\phi_2=c_2$.  This is
analagous  to
defining
  the areas of  surfaces bounded by loops as discussed above.  $V$
does not
commute
   with the $T^N$ but does (trivially) with $Q$ and $A$.

  \section{3+1 Gravity}
The models discussed in the previous section may be converted into
3+1 general  relativity with massless scalar fields by the addition
of  the
 Hamiltonian constraint.  This  constraint is \cite{matter}
 \begin{equation}
  H= \epsilon^{ijk}F_{ab}^iE^{aj}E^{bk} +
E^{ai}E^{bi}\partial_a\phi_n\partial_b\phi_n   -  \Pi^2_n
 \end{equation}
 Such a generalization also involves complexifying
 $A_a$, together with the accompanying reality conditions.

The observables given in the last section do not  Poisson commute
 with this constraint. However we note that the functional $T^0$ is
an
 approximate
solution to the Hamilton-Jacobi equations for gravity determined from
the
constraints
by  replacing the  momenta $E$  and $\Pi$  by $\delta {\cal
S}[A,\phi]/ \delta
A$
 and $\delta {\cal S}[A,\phi]/ \delta\phi $. With ${\cal
S}[A,\phi]\equiv T^0$
we find
 \begin{equation}
 D_a{\delta {\cal S}[A,\phi]\over  \delta A_a^i} =0
 \end{equation}
 \begin{equation}
  F_{ab}{\delta {\cal S}[A,\phi]\over  \delta A_b^i}   +
 \partial_a \phi_n {\delta {\cal S}[A,\phi]\over \delta\phi_n} = 0
 \end{equation}
Equations (3.2-3.3) show that the Hamilton-Jacobi equations for the
model
  discussed here can be solved.
  If we now include the Hamiltonian constraint $H$ however, we have
\begin{equation}
  \epsilon^{ijk}F_{ab}^i{\delta {\cal S}[A,\phi]\over  \delta A_a^j}
 {\delta {\cal S}[A,\phi]\over  \delta A_b^k} =0,
 \end{equation}
 \begin{equation}
 \partial_a\phi_n\partial_b\phi_n {\delta {\cal S}[A,\phi]\over
\delta A_a^i}
 {\delta {\cal S}[A,\phi]\over  \delta A_b^i}=0
 \end{equation}
 but
 \begin{equation}
  {\delta {\cal S}[A,\phi]\over \delta\phi_n} =
 \int_{\gamma[\phi_1,\phi_2]} ds \dot{\gamma}^a(s)
{\delta\gamma^b\over \delta
\phi_n}(s)
Tr[F_{ab}(\gamma(s))U(s,s)].\end{equation}
((3.3) follows since the $T^0$ is diffeomorphism invariant. For $H$,
we note that
each functional derivative $(\delta {\cal S} / \delta A) $ brings
down a term
proportional
to the  tangent vector to the loop, which by definition of the loop
is
orthogonal
to  $\partial_a\phi_n$.  For the first term in $H$, the two tangent
vectors contracted with $F_{ab}$ give zero).

  The fact that the holonomy is a solution of the
Hamilton-Jacobi equation associated with
the Hamiltonian constraint without matter (3.4) has already been
noted
\cite{ash2}.
Here we see that when two scalar fields are included, $T^0$ is a
solution of {\it all} the Hamilton-Jacobi equations if the momentum
term
$\Pi^2$ in $H$ is ignored.
 It may be possible
 to develop a perturbation series that allows the construction of
${\cal S}$ to  better approximations. A perturbation series in powers
of the gradient of the scalar field for the ADM variables and scalar
fields has
been considered in \cite{sal} and the methods there may be useful in
the
present
 context as well.

 For the purpose of addressing the integrability of the model
 discussed here, the solution of the Hamilton-Jacobi equations
presented
 above are not sufficient since there are an insufficient number
  of integration momenta in the solution. A more general method
  of solving the H-J equations for the diffeomorphism and
  Gauss constraints without matter has been
  discussed by Newmann and Rovelli \cite{newc}, and in the  following
   we note their results together with an  extension
  to include the matter fields: the Gauss law
  equation (3.2) is solved by
  \begin{equation}
  S[A;U^{(a)},V^{(a)}] = \sum_{(a)}\int du^{(a)}\int dv^{(a)}
  T^0[A,U^{(a)},V^{(a)}](u^{(a)},v^{(a)})
  \end{equation}
 where $U^{(a)},V^{(a)}$ are three pairs of scalar fields $(a=1,2,3)$
 parametrizing the solution, and the $T^0$ is constructed as in
 (2.7), but with loops obtained from the intersection of the surfaces
  $U^{(a)}(x)=u^{(a)}$, $V^{(a)}(x)=v^{(a)}$. The part of the
  diffeomorphism constraint involving the gravitational variables
   $(A,E)$ can be written in an intuitively understood form
   involving the $U^{(a)},V^{(a)}$ and the
  coordinates $Q_U^{(a)}, Q_V^{(a)}$ conjugate to the
  {\it  reduced momenta}, which are  determined
  by the three congruences associated with $U^{(a)},V^{(a)}$
  \cite{newc}.
  Using this, the reduced diffeomorphism constraint becomes
  \begin{equation}
   C_a = \Pi_n\partial_a \phi_n + Q_U^{(a)}\partial_a U^{(a)}
 + Q_V^{(a)}\partial_a V^{(a)}
  \end{equation}
 The H-J equation associated with this constraint
is  obtained by setting $\Pi_n = (\delta S/\delta\phi_n)$ and
$Q_U^{(a)}=\delta S/\delta U^{(a)}$, $Q_V^{(a)}=\delta S/\delta
V^{(a)}$.
This has a solution
\begin{equation}
S[U^{(a)},V^{(a)},\phi_n; P^{(a)},p_n]
= \sum_{(a)}\int d^3x
\tilde{V}U^{(a)}(x)P^{(a)}(V^{(a)}(x))\phi_n(x)p_n(V^{(a)}(x))
\end{equation}
where the $P^{(a)},p_n$ are $n+3$ integration momenta parametrizing
the
solution and the density
$\tilde{V}=
\epsilon^{abc}\partial_aV^{(1)} \partial_bV^{(2)}\partial_cV^{(3)}$.
$S$ is clearly diffeomorphism invariant since its integrand is a
density constructed solely from the configuration variables
$U^{(a)},V^{(a)},\phi_n$.
This solution is the generalization of the results of
[15] to include matter. An extension of this result to general
relativity would require rewriting the Hamiltonian constraint
 in terms of the reduced momenta $P^{(a)},p_n$ and their
 conjugate coordinates, and then solving the
 associated H-J equation. To address the question of
 integrability, such a solution should involve $n+2$ integration
momenta.

For  general relativity there is also an issue as to what phase space
functionals should
be called physical observables since the Hamiltonian constraint
generates
 {\it both} time reparametrizations {\it and} evolution of the
phase space variables from one spatial surface to another.
In the sense of physical
observables as those that commute with all the constraints but $H$,
the $T^N$ defined above  form a kinematically gauge invariant  set
whose classical evolution via $H$
may be  studied.

\section{Quantization}
We can study the quantization of this model in the connection/scalar
field
representation or  the loop representation. These are discussed and
compared
 in this section.

 (i) {\it Configuration representation}.
 For  Dirac quantization, we can write down
a diffeomorphism and gauge invariant wavefunction. This is the Wilson
loop $\Psi[A,\phi_1,\phi_2](c_1,c_2) = T^0$.  These Wilson loop
states
are therefore
{\it physical states} of the quantum theory for the action $S$. These
are
 a  two parameter $(c_1,c_2)$ family of states and so most likely
  do not make up the full state space. (This point is discussed
further
   below).

For  gravity with scalar fields, $\Psi[A,\phi_1,\phi_2](c_1,c_2)$
  is not a full
solution to the Hamiltonian constraint. It is annihilated by
{\it all} terms in the
Hamiltonian constraint {\it except} for the momentum squared terms of
the scalar field.
(The reasons are the same as  those discussed above for the
Hamilton-Jacobi
 equations).
Thus one may  perhaps view these states as solutions to the entire
set
 of  Dirac quantization conditions in the approximation
 that the scalar field momenta are constants.
 Since the states $\Psi$ are diffeomorphism invariant in the $A,\phi$
representation
one can attempt to complete the quantization in this representation
by some
type of perturbation expansion to incorporate the momenta of the
scalar fields.

(ii) {\it Loop representation}.
For pure gravity or the model given by (2.1),
an alternative representation for the quantum theory
can  be obtained by converting the closed Poisson algebra of the
(Gauss law invariant) loop observables into an operator algebra
on functions of loops \cite{lc1}. In the limit $\hbar\rightarrow 0$,
this operator algebra reduces to the classical Poisson algebra.
The motivation for this representation (for pure gravity) is that,
unlike the configuration representation, it allows
 the solution of the diffeomorphism constraints in a
 natural way: The diffeomorphism invariant information
 in a loop are its knot invariants and so the
  `quantum numbers' labelling the  physical states are the knot
invariants.

 In the present context with matter fields, although some solutions
of the  diffeomorphism constraint can be obtained in the
configuration
 representation, one might attempt the same with the
observables $T^N$, but there is a potential problem:
 If the scalar field momenta are included as part of the observables
to be represented as operators in the quantum theory, then the
Poisson algebra of the $T^N$ doesn't close unless it is extended
 as discussed previously. Such an extension is inelegant but
 can be done, and one can then attempt to construct a loop
 representation analagous to the one for pure gravity. This
 is under study.

  One can on the other hand proceed with  a `reduced' quantization
 of the model where the scalar field momenta are not represented
 as operators, in which case the {\it same} loop representation
 as that for gravity may be used. This is necessarily limiting
 since all the basic variables of the classical theory
 will not be realized as operators in the quantum theory.
 One can nonetheless see what can be learned of the quantum theory
 from this (restricted) quantization.
 (In this regard Van Hove's result \cite{vhove} is worth pointing
out:
 It is not possible to convert {\it all} the fundamental variables
  (satisfying some closed Poisson algebra) into operators
  such that the  correspondence is maintained between all
  the Poisson brackets and commutators). Proceeding in this manner
 would give a loop representation on which there would be no need to
 impose the diffeomorphism constraints since the observables are
  already invariant. The physical states would therefore no longer
  be labelled by knot classes, but still by the same two parameter
set.

An alternative approach to the loop representation when there is
 matter is
 to not work with the $T^N$ made from matter loops,
  but rather to have auxilliary loops parametrizing the gravitational
  loop observables. This would lead to the usual loop space
representation
  for the gravitational degrees of freedom, with the physical
  states labelled by knot invariants. The matter would be
incorporated
   seperately resulting in product states of the form
   $|knot class>|matter>$ \cite{csurf}.

 {\it Comparison}. We have seen that there appear to be
 two different approaches to incorporating matter in the loop
 representation: (1) Work with a (suitably completed or
 reduced) set of diffeomorphism invariant matter-loop observables
 and find a representation of their Poisson algebra to obtain
the quantum theory, or (2) treat loops as auxilliary and
attempt to obtain suitably defined product states for the
matter gravity variables. The latter would give a larger
set of physical states because a general knot is
not obtainable from the configurations of two scalar
 fields discussed here.
 This  shows that the two parameter set of states given above
 are most likely not all the physical states.
  (In this regard it is worth emphasizing that in general, the chosen
representation  determines the size of the Dirac quantization
state space, and in the absence of an innerproduct
the sizes of the state spaces in different representations
may not be the same \cite{kuchgrg}).

The representation of the gravitational part of the functionals
$T^0,Q,V,A$  as operators on the loop representation space has been
studied in ref. \cite{ars}.  Since  the  definitions of these
observables
here  differ only in the replacement of auxiliary variables
(loops and surfaces)
by configurations of the two scalar fields,
the gravitational part
of the representation needs to be extended to include a
representation
for the scalar fields.
 One such extension has been discussed in \cite{csurf} and a similar
 procedure may be applied
 here. The essential results \cite{ars} concerning the quantization
 of the spectra of $V$ and
 $A$ remains unaffected.  The new features are that the scalar fields
allow
 diffemorphism invariant definitions for the areas of  surfaces with
boundary
 and the volumes of particular regions bounded by surfaces.

 \section{Discussion}
 We have given a four dimensional generally covariant theory that has
 an infinte number of
 observables (which are the constants of motion). In particular
  since the Hamilton-Jacobi equations can be
  solved, the theory may be completely integrable if the functionals
  relating the new and old phase space variables can be inverted.

 The Dirac quantization  conditions for the theory can also be
 solved exactly to give a class of  physical states in the coordinate
representation - the physical states are traces of holonomies,
with the loops  determined by scalar fields.
It is possible to construct a reduced  loop space
 representation for the observable algebra
 by choosing not to represent the scalar field momentum. On the other
hand
 one can also attempt to complete the classical loop variables by
extending
 the algebra of the $T^N$, and then seeking a loop space
representation.
  The model also has a number of other observables which are known to
  have well-defined operator versions on the loop representation
space
 \cite{ars}.

 The inclusion of the Hamiltonian constraint  on the phase space of
the
 model  gives general relativity  with scalar fields and so the above
results
 partially carry over to this case. The observables are now gauge
invariant
only  under  transformations generated by the kinematical
constraints, and
one obtains approximate solutions of the Dirac quantization
conditions and
the additional Hamilton-Jacobi equation.

 The role of the scalar fields is similar in a way to
   choosing coordinates in general relativity  that
are based on the presence of fluids whereby the fluid particles mark
space points and their clocks specify the time foliation.  Such
`reference
fluids' \cite{kuchtor}
have been used recently to attempt to solve the problem of time  in
quantum
gravity.
In this approach coordinate conditions are chosen in the action via
lagrange
multipliers
thereby  breaking diffeomorphism invariance, and then the theory is
re-parametrized
to restore this invariance. The resulting theory gives an effective
source
term for the Einstein equations  with the source determined by the
coordinate
conditions.
    In this spirit the scalar fields here may be viewed as the
sources that
specify matter
    loops on which the diffeomorphism invariant loop observables are
based. The
 loop observables do not commute with the Hamiltonian constraint but
 we still have an interesting set of  spatial diffeomorphism
invariant
 variables  that form a closed algebra whose dynamics  may be worth
studying.
The   physical picture seems rather interesting since as  the scalar
fields
evolve
 one can envision the matter loops joining and  breaking.

An extension of the results may be to see if  diffeomorphism
 invariant loop observables can be found by
 coupling Yang-Mills and spinor fields to the model  given here, and
to general
relativity.
 Such observables, with  loops constructed out of the different
matter fields,
may  provide one
 way to include general matter fields into the loop representation.

 \acknowledgements
 I thank Andre Barvinsky and Carlo Rovelli for interesting
discussions.


\begin{references}
\bibitem{ash1}  A. Ashtekar, {\it Non-perturbative canonical
gravity}, Lecture
notes
in collaboration with R. S. Tate, World Scientific (1991).

\bibitem{lc1} C. Rovelli and L. Smolin, Phys. Rev. Lett. 61, 1155
(1988); Nucl.
Phys.
B331, 80 (1990).

\bibitem{dw} B. S. DeWitt, in {\it Gravitation, an intorduction to
current
research}, ed.
L. Witten  (Wiley, NY 1962).

\bibitem{rov} C. Rovelli, Class. and Quant. Gravity 8 (1991) 297,
317.

\bibitem{kuchtor} K. V. Kuchar and C. G. Torre,  Phys. Rev. D43
(1991) 419; D44
(1991) 3116.

\bibitem{3grav} A. Ashtekar, V. Husain, C. Rovelli, J. Samuel, and L.
Smolin,
Class. Quantum Grav.  6,  L185 (1989).

\bibitem{ars} A. Ashtekar, C. Rovelli, and L. Smolin, Phys. Rev.
Lett. 69, 237
(1992).

 \bibitem{csurf} C. Rovelli, {\it A generally covariant quantum field
theory},
University  of
 Pittsburgh preprint (1992).

 \bibitem{lsurf} L. Smolin, {\it Quantum theory of surfaces},
Syracuse
University
preprint
  (1992); {\it Time, measurement and information loss in quantum
cosmology}, to
  appear in the D. Brill festshrift.

\bibitem{hk} V. Husain and K. V. Kuchar, Phys. Rev. D42, 4070 (1990).

\bibitem{h} V. Husain, Class. Quantum Grav. 9, L33 (1992).

\bibitem{ash2} A. Ashtekar. {\it New perspectives in canonical
gravity},
Bibliopolis,
Naples(1988).

\bibitem{sal}  D. S. Salopek and J. M. Stewart, Class. Quantum Grav.
9, 1943
(1992).

\bibitem{matter} A. Ashtekar, J. Romano, and R. S. Tate, Phys. Rev
D40 9 (1989)
2572.

 \bibitem{newc} E. T. Newman and C. Rovelli, Phys. Rev. Lett. 69,
(1992)
  1300.

 \bibitem{kuchgrg} K. V. Kuchar, in {\it Proceedings of the 13th.
 International Conference  on General Relativity and Gravitation},
 ed. C. Kozameh (IOP publishing, Bristol 1992).

 \bibitem{vhove} L. Van Hove, Acad. Roy. Belg. Bull. Cl. Sc. 37
(1951)
 610.
 \end{references}
 \end{document}